\documentclass[10pt]{wlscirep}
\usepackage{amsmath}
\usepackage{amssymb}
\usepackage{verbatim}
\usepackage{fancyhdr}
\usepackage{array}
\usepackage{subfigure}
\usepackage{textcomp}
\usepackage{gensymb}
\usepackage{graphicx}
\usepackage{color}
\usepackage{soul}
\usepackage{bm}
\usepackage[utf8]{inputenc}
\usepackage[T1]{fontenc}
\usepackage{xcolor}

\title{Be Water, My Antennas: Riding on Radio Wave Fluctuation in Nature for Spatial Multiplexing using Programmable Meta-Fluid Antenna}

\author[1,2]{Baiyang Liu (IEEE Senior Member)}
\author[3,4]{Kin-Fai Tong (IEEE Fellow)}
\author[4,5,*]{Kai-Kit Wong (IEEE Fellow)}
\author[5]{Chan-Byoung Chae (IEEE Fellow)}
\author[2,*]{Hang Wong (IEEE Fellow)}

\affil[1]{College of Big Data and Internet, Shenzhen Technology University, Shenzhen, 518118, China}
\affil[2]{State Key Laboratory of Terahertz and Millimeter Waves, Department of Electrical Engineering, City University of Hong Kong, Hong Kong SAR, China}

\affil[3]{School of Science and Technology, Hong Kong Metropolitan University, Hong Kong SAR, China}

\affil[4]{Department of Electronic and Electrical Engineering, University College London, London, WC1E7JE, U.K.}

\affil[5]{Yonsei Frontier Lab, Yonsei University, Korea}
\affil[*]{Corresponding authors: Kai-Kit Wong (kai-kit.wong@ucl.ac.uk) and Hang Wong (hang.wong@cityu.edu.hk)}

\begin{abstract}
Interference and scattering, often deemed undesirable, are inevitable in wireless communications, especially when the current mobile networks and upcoming sixth generation (6G) have turned into ultra-dense networks. Current approaches relying on multiple-input multiple-output (MIMO) combined with artificial-intelligence-aided (AI) signal processing have drawbacks of being power-hungry and requiring wide bandwidth that raise scalability concerns. In this article, we take a radical approach and utilize the channel fading phenomenon to our advantage. Specifically, we propose a novel meta-fluid antenna architecture, referred to as the `fluid' antenna system (FAS), that can freely surf on radio wave fluctuations, like `fluid' figuratively speaking, with fine resolution in space to opportunistically avoid interference, eliminating the need for expensive signal processing. Our experimental results demonstrate that under rich scattering conditions, the proposed meta-fluidic architecture is able to exploit the natural ups and downs of radio waves in space for spatial multiplexing. These breakthrough results show that scattering can be desirable not harmful and interference can be dodged not suppressed, fundamentally changing our perception of fading and our understanding on how interference should be managed in wireless communications networks.
\end{abstract}

\begin{document}

\flushbottom
\maketitle

\section*{Introduction}
Although the fifth-generation (5G) mobile systems have been painted to be a revolutionary technology, it is yet to impress average smartphone users, as the catchy use cases namely holographic communications \cite{el2022holographic, zhang20236g} and tactile Internet \cite{fettweis2014tactile, simsek20165g}, demonstrated in 5G promotions, are nowhere to be seen so far. This is because in promotional events, dedicated spectral and energy resources are ensured to enable the delivery of these use cases but in reality, resources are limited and mass provision of such highly demanding services is inconceivable using today's technologies. This has urged communication practitioners, researchers and scientists to ponder on the improvements required for the next generation, a.k.a.~sixth-generation (6G) \cite{jia2023valley, dang2020should,tariq2020speculative}. Among other ambitious key performance indicators (KPIs), one standout target is to hit more than $1000~{\rm bps/Hz}$ spectral efficiency in 6G \cite{you2023toward}.\footnote{Spectral efficiency is the measure of the rate achievable in bits per second for each Hertz of bandwidth provided, over a given channel. The classical expression for spectral efficiency, according to Shannon \cite{shannon1948mathematical}, is $\log_2(1+{\rm SNR})$ where ${\rm SNR}$ is the received signal-to-noise ratio (SNR). In situations where $U$ users share the same physical channel, the expression becomes $U\log_2(1+{\rm SINR})$ where ${\rm SINR}$ denotes the per-user signal-to-interference plus noise ratio (SINR) if every user is assumed to have identical performance. Evidently, the above discussion has assumed that each user has only a single antenna. But if each user is given a multiple-input multiple-output (MIMO) channel, each user's spectral efficiency will become $r\log_2(1+{\rm SNR})$ where $r=\min(n_T,n_R)$ and $n_T$ and $n_R$ are, respectively, the number of antennas at the transmitter and receiver. However, $n_R$ is usually small due to space and cost consideration and $r$ will be small in practice. Therefore, with or without MIMO, if the network needs to achieve more than $1000~{\rm bps/Hz}$ spectral efficiency, then the conclusion remains the same that spectrum sharing over a large number of users will be the way forward.} This target would mean that the communication link needs to operate at an out-of-this-world SNR of $3000~{\rm dB}$ if a single-user point-to-point system is considered. More accurately, a sensible interpretation of the target is to require the network to share the spectrum by a large number of users and ensure that each user  still has acceptable performance.

Multiple access techniques \cite{xiong2023breaking, bozinovic2013terabit} govern how users are given access to the available spectral resources for communications. Traditional approaches rely on time-division multiple-access (TDMA) and frequency-division multiple-access (FDMA) where users are provided with orthogonal, non-overlapped channels for communications because of their simplicity in interference management. However, as mobile networks attempt to catch up with the rising demands under limited spectrum, it is increasingly necessary to share the same time-frequency channel over multiple communications. In doing so, proper interference management needs to be in place to ensure reliable communications. In 5G New Radio (NR), this is being achieved by the massive multiuser multiple-input multiple-output (MU-MIMO) technology \cite{urquiza2022spectral}. 

The principle of MU-MIMO is simple. Assuming the availability of the channel state information (CSI)\footnote{CSI corresponds to the information regarding the amplitude and phase shift imposed onto the received signal from the wireless channel, which can also include the angle-of-departure (AoD) and/or angle-of-arrival (AoA) of each propagation path when appropriate.} of the users, a multi-antenna base station (BS) can utilize signal processing to mix the signals cleverly from its antennas before transmitting to the users---a method called MU-MIMO precoding. 

Unfortunately, the experience of massive MU-MIMO in 5G-NR has been somewhat underwhelming, raising doubts about the potential of extra-large MIMO (XL-MIMO) for 6G \cite{wang2023extremely}. The main challenge is scalability, particularly the difficulty in upscaling the CSI acquisition process with many users and the complexity of computing the precoding solution. Reliable CSI acquisition is crucial for effective MU-MIMO precoding, but wireless channels change rapidly, making it harder to maintain CSI accuracy with more users. It is also worth mentioning two emerging multiple access techniques, non-orthogonal multiple access (NOMA) \cite{dai2015non} and rate-splitting multiple access (RSMA) \cite{mao2022rate}, which propose overlapping users on the same physical channel using advanced coding and decoding schemes. While NOMA was seriously considered during the 5G development cycle \cite{chen2018toward}, RSMA is now entering 3GPP discussions for beyond 5G and 6G standardization. However, both NOMA and RSMA face challenges with CSI feedback and high decoding complexity, limiting their potential as massive spectrum-sharing solutions. A fundamentally different approach is needed to complement existing and emerging MU-MIMO technologies.

With the rapid growth of artificial intelligence (AI), some might feel optimistic about addressing the scalability issue in MU-MIMO \cite{letaief2019roadmap}. As a matter of fact, deep learning has already been utilized to reduce the CSI feedback for massive MIMO systems in 5G \cite{wen2018deep}. Deep learning is also anticipated to be applicable to a wide range of design problems in the physical layer of 6G \cite{he2019model}. With that being said, however, the system performance is fundamentally dictated by the physical signals at hand, and AI will be powerless to change that. In light of this, the objective of this work is to break the physical boundaries of existing communication systems and explore opportunities to innovate multiple access technologies for wireless communications that scales well with the number of users. 

In order to rethink multiple access, the concept of a new form of reconfigurable antennas, referred to as `fluid' antenna system (FAS), which has recently been introduced to become an enabling technology for 6G \cite{wong2020performance, wong2020fluid}, is highly appealing. FAS represents any software-controllable shape-flexible position-flexible antenna system for wireless communications. FAS is first of its kind in the wider area of reconfigurable antennas. From the information-theoretic perspective, FAS can be viewed as some new degree-of-freedom in the physical layer for enhancing communication performance. A recent tutorial article \cite{new2024tutorial} provides a comprehensive coverage on many topics on FAS. Remarkably, through position reconfigurability, FAS gives a transceiver\footnote{This can be a BS, an access point, handset, tablet or any Internet-of-Things (IoT) device with wireless connection capability.} the ability to access the received signals in fine space. In contrast to having only one received signal at a {\em point} in space from a conventional fixed-position antenna, FAS makes it possible to receive signals from a prescribed {\em spatial region} with fine resolution. This change is fundamental because this expands the dimensionality of the received signals without necessarily increasing the number of radio-frequency (RF) chains accordingly. Importantly, this capability enables access to the ups and downs of the wireless channel in space, which can be exploited for multiple access. In particular, a wireless channel, usually modelled as a complex coefficient characterizing both magnitude attenuation and phase shift, sums up the overall effect of radio wave propagation in a wireless environment and this can also be interpreted as the superposition of multiple propagation paths between the transmitter and receiver over a wireless link. A classical phenomenon is that a wireless channel imposes random attenuation and phase shift and the channel can disappear when the multiple paths happen to cancel each other, which is referred to as {\em fading}.\footnote{It is worth saying that fading is widely viewed as the `curse' for wireless communications and decades of efforts have been spent to mitigate this to restore some stability for communications and attempt to achieve a decent reception performance.} This channel fading coefficient changes if the position of reception is different, as different propagation paths mix differently. Now, this is where FAS can be uniquely effective. Specifically, FAS is able to observe how the wireless channel varies over a given spatial region with fine resolution, and the fading phenomenon means that the FAS will likely see that  at some positions, the channel disappears. In multiuser scenarios, FAS therefore can choose to activate the position where the channel of the aggregate interference disappears (or becomes very weak) and this is the fluid antenna multiple access (FAMA) concept proposed in \cite{wong2023slow}. The exciting result here is that interference avoidance is achieved without the need of CSI at the BS as in MU-MIMO precoding nor advanced coding and decoding as in NOMA and RSMA. Our interpretation of FAMA is to ride on the radio wave fluctuation and exploit the opportunity created by nature for spatial multiplexing,\footnote{Spatial multiplexing refers to multiple access techniques that separate users in the spatial domain.} resembling Bruce Lee's Jee Kune Do's famous philosophy on combat {\em ``Be Water, My Friend''}. While Bruce Lee approaches fighting by imitating water for adaptivity, FAMA approaches multiple access by the same idea {\em ``Be Water, My Antennae''}.

The concept of FAMA was introduced by Wong {\em et al.}~in 2022 \cite{wong2023slow, wong2021fluid} while subsequent efforts further looked to address its challenges in signal processing \cite{waqar2023deep} and combine this with other technologies \cite{waqar2024opportunistic}. Nevertheless, existing results are largely limited to theoretical studies, and how FAMA can be realized in practice is not well understood. To implement a FAS, a straightforward approach is to employ liquid antennas, due to its shape flexibility \cite{huang2021liquid,dash2023selection}. Recently in \cite{shen2024design}, liquid-metal antennas were adopted to design a position-flexible FAS and the measurement data obtained were also used to test the feasibility of FAMA. Despite the positive results, the major criticism of the liquid-based approach is the slow response time in changing the position of FAS. The required response time is in the millisecond range or less (i.e., the channel coherence time) but changing the position of a liquid-based radiating object takes a little less than a second in the best case. A promising alternative is to use reconfigurable pixels to realize FAS \cite{song2013efficient}. In this approach, a reconfigurable radiating surface is composed of a matrix of tiny pixels, with the connections between pixels being optimized to reconfigure the radiation characteristics. This approach has recently been explored to implement FAS in \cite{zhang2024pixel}, producing encouraging results. However, the results in \cite{zhang2024pixel} are still in a very early stage and there is lack of understanding of how the optimized connections between pixels link to the specific features in radiation. Also, whether the FAS in \cite{zhang2024pixel} is effective for FAMA is yet to be seen. There are also the ideas of coding on metasurfaces \cite{wang2024multichannel, zhang2018space,fusco2021single,zhang2021wireless,wu2022sideband,wang2023manipulations,wu2023universal, wu2024synthetic}, leaky-wave antennas \cite{karl2015frequency, memarian2015dirac, matsumoto2020integrated, xu2023arbitrary}, pixel-based antenna \cite{zhang2024pixel} and moveable antenna \cite{zhu2024historical} that are applicable to the design of FAS even though there have been no attempts except \cite{zhang2024pixel} so far to utilize these techniques to realize the concept of FAS. Presumably, it is possible to use positive-intrinsic-negative (PIN) diodes to control the operating state of each metasurface element, and the theory of leaky-wave antennas can be useful to design substrate integrated waveguides (SIWs) that manage the propagation of radio waves.

\begin{figure}[!h]
\captionsetup{justification=justified, singlelinecheck=false}
\centering\includegraphics[width=17cm]{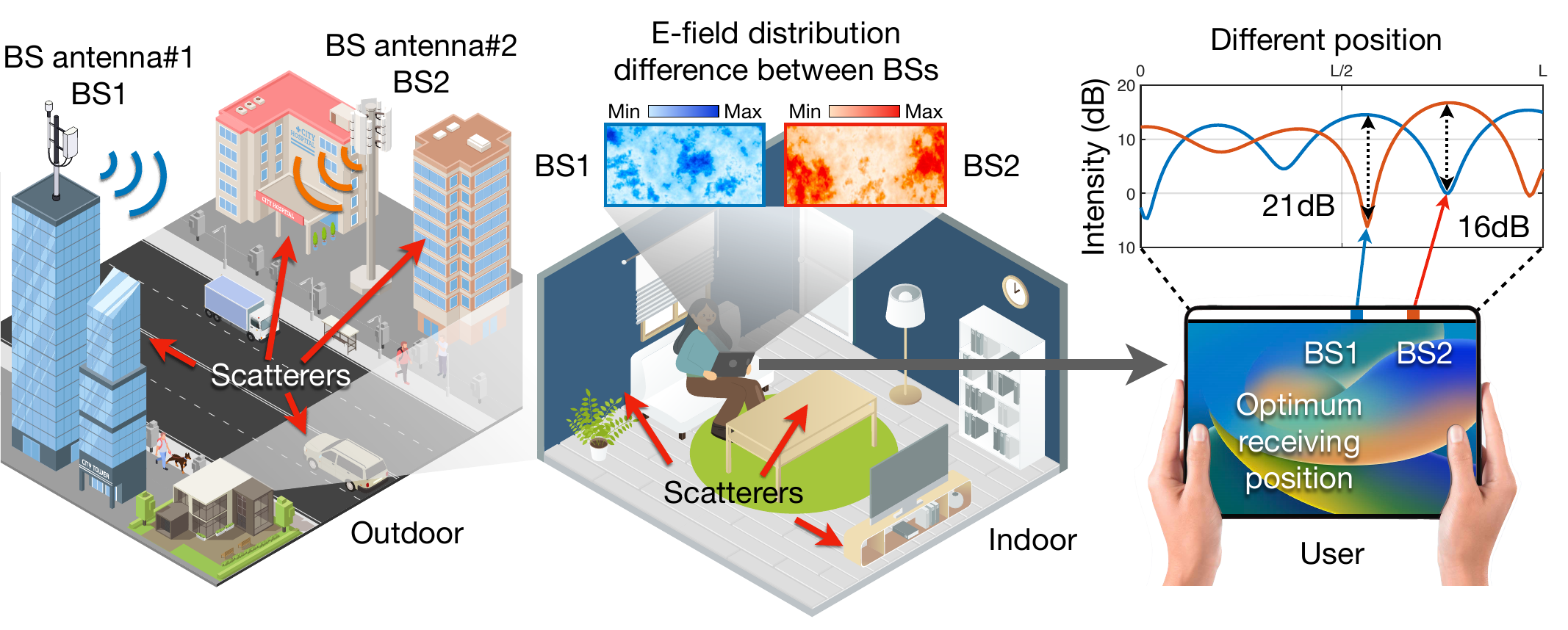}
\caption{\textbf{Illustration of the concept of FAS and FAMA.} A downlink communication system is shown, where two distributed BSs are responsible for transmitting data to two mobile users in a typical rich-scattering environment (both outdoor and indoor). The figure in the middle also shows a user with a FAS mounted on her laptop, enabling the observation of the E-field distributions of the signals arriving from the two respective BSs. The figure on the right then illustrates the concept of FAMA. Specifically, this figure focuses on a particular horizontal dimension of the FAS and displays how the intensity of the radio signals changes along this dimension. It is clear that the radio signal intensity fluctuates, which gives rise to the opportunity for the FAS to avoid interference.}\label{conceptual_illustration}
\end{figure}

In this work, we take a radical approach that views scattering as a desirable factor that provides opportunity by nature to harness interference, as opposed to negating scattering by heavy signal processing to restore channel stability. This is achieved by a new form of reconfigurable antenna, referred to as meta-fluid antenna by integrating leaky-wave antennas with coding-on-metasurface, to exploit the fine resolution in space to utilize the interference null naturally created due to scattering. The proposed idea follows the spirit of Bruce Lee's philosophy in martial art, {\em ``Be Water, My Friend''} and is enabled by the novel fully electronically programmable meta-fluidic architecture. Here we present the first demonstration of FAMA and show the potential application scenarios, which incorporates artificial massive-rich scattering transmitters and introduces a novel receiver called the meta-fluid antenna (see Figure \ref{conceptual_illustration}). We have successfully harnessed the power of electromagnetic (EM) wave dynamics in space and time by strategically activating the elements of the metasurface to the position in which the SINR is maximized for multiple access, preventing the need for expensive signal processing at the BS side. This new approach has far-reaching implications for a broad range of wireless applications, sensing networks and encryption technologies.

\section*{Results}
\subsection*{Principles of FAMA}
Consider a wireless communication network where $U$ users are being served on a shared physical channel by a BS in the downlink.\footnote{This model is motivated by its generality and the need for aggressive spectrum sharing in 5G/6G.} The BS has $U$ fixed-position antennas, with each antenna dedicated to transmit communication signals to one of the users. The assignment of BS antenna to user is unimportant in this work but can be easily done when needed in practice. At the user side, each user is equipped with a `fluid'-like antenna (referred to as FAS) that is considered to have the ability to ride on the radio waves and switch to a `good' radiating position in accordance with the changing wireless channel conditions. The FAS at each user is assumed to have $N$ preset positional points, arranged in a two-dimensional (2D) grid of $I$ rows and $J$ columns over a given surface; hence $N = I \times J$. All of these selectable positions share a common RF chain. Under this model, the received signal at the $(i,j)$-th position of FAS for user $u$ can be expressed as: \begin{equation}\label{FAMA model}
r_{i,j}^{(u)}=s_u g_{i,j}^{(u, u)}+\sum_{\tilde{u}=1 \atop u \neq u}^U s_{\tilde{u}} g_{i,j}^{(\tilde{u}, u)}+\eta_{i,j}^{(u)},
\end{equation}
where $u\in \{1,2,\dots,U\}$, $i\in \{1,2,\dots,I\}$, $j\in \{1,2,\dots,J\}$ and \( s_u \) represents the transmitted symbol intended for user \( u \), while \( \eta_{i,j}^{(u)} \) denotes the zero-mean complex additive white Gaussian noise (AWGN) at the \( ({i,j}) \)-th position of the user. The term \( g_{i,j}^{(\tilde{u},u)} \) describes the fading channel from the $\tilde{u}$-th BS antenna to the \( ({i,j}) \)-th position of user \( u \). When $\tilde{u}\ne u$, this corresponds to the interfering channel while the $\tilde{u}=u$ case gives the desired channel. We will focus on the timescale where the fading channels are regarded as static to ease our discussion without loss of generality. In practice, channels do change however but the same process can be repeated every time changes happen.

In FAMA, user $u$ switches its FAS to the position that maximizes the SINR, or signal-to-interference ratio (SIR) when noise is negligible at high SNR \cite{wong2023slow}. That is, the optimal position $\left.{(i^*,j^*)}\right|_u $ is selected by
\begin{equation}
\left.{(i^*,j^*)}\right|_u =\arg \max _{i,j} \frac{\mathbb{E}\left[\left|g_{i,j}^{(u, u)} s_u\right|^2\right]}{\mathbb{E}\left[\left|\sum_{\tilde{u} = 1 \atop
\tilde{u} \neq u}^U g_{i,j}^{(\tilde{u}, u)} s_{\tilde{u}}+\eta_{i,j}^{(u)}\right|^2\right]}
\approx\arg \max _{i,j} \frac{\left|g_{i,j}^{(u, u)}\right|^2}{\sum_{\tilde{u}=1 \atop \tilde{u} \neq u}^U\left|g_{i,j}^{(\tilde{u}, u)}\right|^2},
\end{equation}
in which $\mathbb{E}[\cdot]$ returns the expectation of an input random entity. The theoretical performance in terms of outage probability for FAMA has been studied in \cite{yang2023performance, xu2024revisiting}, revealing promising results in multiple access.

What is missing in the literature is the experimental validation of the effectiveness of FAMA. This does not only require the highly reconfigurable FAS hardware to be designed and fabricated but also the experiments that give measurement data to validate the performance in real environments. This will be the goal of this paper.

\subsection*{Meta-fluid antenna}
\begin{figure}[!h]
\captionsetup{justification=justified, singlelinecheck=false}
\centering\includegraphics[width=16cm]{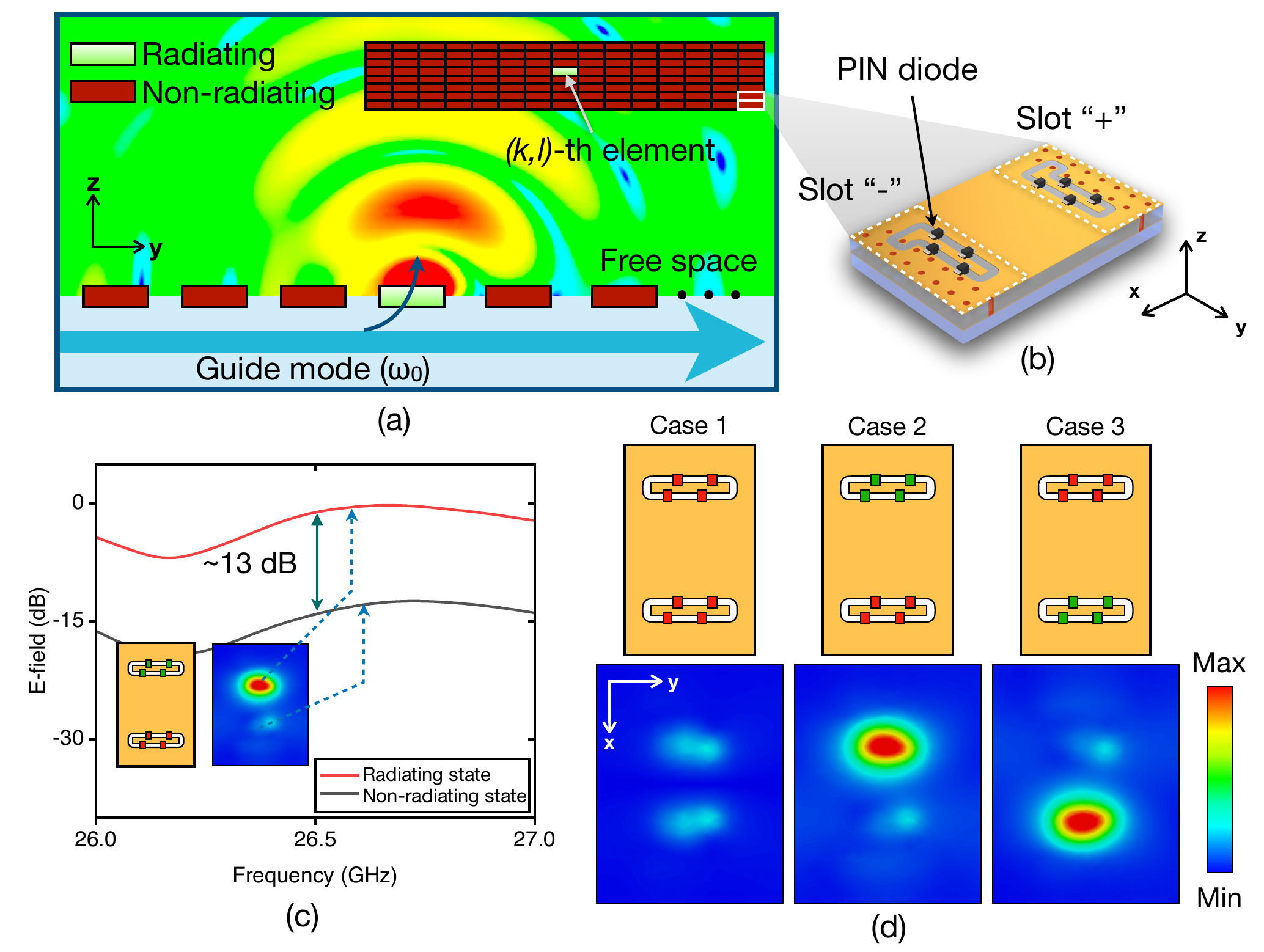}
\caption{\textbf{The working principle of the proposed meta-fluid antenna.} (a) This shows a schematic representation of the meta-fluid antenna architecture with $8\times15$ radiating elements and the simulated radiating pattern in the $yz$-plane if a given element is activated. (b) This figure shows the basic 3-dimensional (3D) structure of a meta-atom which consists of an upper element and a lower element. Each element is connected and controlled by 4 PIN diodes. (c) This figure illustrates the simulated E-field results against frequency focusing on one meta-atom if the upper element is in an `ON' state but the lower element is in an `OFF' state. The results show an approximate $13~{\rm dB}$ difference in their E-fields between the two elements. (d) The figures display the E-field distributions in the $xy$-plane in the three possible states for the meta-atom.}\label{metafluid_antenna}

\end{figure}

Our proposed meta-fluid antenna consists of a 2D array of $N=120$ elements, arranged in $I=8$ rows and $J=15$ columns. The elements can be dynamically switched between radiating and non-radiating states, as illustrated in Figure \ref{metafluid_antenna}(a). This architecture effectively integrates $120$ individual positions within a single RF chain. By utilizing a field-programmable gate array (FPGA) controller, a single element with $4$ PIN diodes can be selectively activated to radiate energy from the waveguide to air, while the remaining elements remain in a non-radiating state. This enables the realization of a meta-fluid, with the interpretation that a genie is riding on the radio waves on the metasurface and able to dynamically activate a particular position for reception.

Technically speaking, a meta-atom is made up of two elements, referred to as the slot `$+$' and slot `$-$' elements, respectively, see Figure \ref{metafluid_antenna}(b). An important distinction between a meta-atom and a selectable position is that each meta-atom includes two selectable positions, one for each slot. Each element functions as a waveguide-fed magnetic dipole, extracting energy from the waveguide and emitting EM waves into the surrounding space. Given that the lattice size of the meta-atom is much smaller than the wavelength, the elements can be considered as sampling points that capture the energy of the guided waves propagating within the waveguide at their respective positions. The magnetic field produced by the $(k,\ell)$-th meta-atom in the far-field region at distance $d_{k,\ell}$ is written as
\begin{equation}\label{eq:1}
{H}_{k,\ell}(\theta)=-\frac{\pi f_0^2{m}_{k,\ell}\cos\theta}{d_{k,\ell}}  e^{-j k_0 d_{k,\ell} \sin \theta},
\end{equation}
where $f_0$ is the carrier frequency, $\theta$ is the observation angle relative to the meta-fluid antenna, $k_0$ is the propagation constant in free space, and $m_{k,\ell}$ is the polarized magnetic dipole moment at time $t$ excited by the meta-atom as
\begin{equation}\label{eq:2}
{m}_{k,\ell}={P}_{k,\ell} H_{k,\ell}={P}_{k,\ell} H_0 e^{-j \xi_{\rm gw} x_{k,\ell}} e^{j 2 \pi f_0 t},
\end{equation}
where \(P_{k,\ell}\) denotes the complex magnetic polarizability at the $(k,\ell)$-th position, and \(H_{k,\ell}\) signifies the instantaneous magnetic field produced at the $(k,\ell)$-th position within the waveguide structure. Additionally, $H_0$ is the corresponding magnitude of the magnetic field, $x_{k,\ell}$ denotes the corresponding propagation distance in the waveguide, and $\xi_{\rm g w}$ is the propagation constant for a guided wave, given by
\begin{equation}\label{eq:3}
\xi_{\rm gw}=\sqrt{\omega^2 \mu \varepsilon-\left(\frac{\pi}{W_{\rm eff}}\right)^2},
\end{equation}
where $\mu$ and $\varepsilon$ are the permeability and permittivity of the medium, $\omega$ is the carrier angular frequency, and $W_{\rm eff}$ denotes the effective width of the waveguide. 

Noting that the magnetic current within the waveguide exhibits a phase reversal between the upper and lower slots, for ease of reference, we refer to them as the slot `$+$' and the slot `$-$', respectively. For differentiation between the two slots along the longitudinal axis of the waveguide, the magnetic polarizability and distance to the far-field observation point for the upper slot are denoted as \(P_{k,\ell}^{+}\) and \(d_{k,\ell}^{+}\), respectively, while those for the lower slot are \(P_{k,\ell}^{-}\) and \(d_{k,\ell}^{-}\). The slots in each meta-atom are controlled by PIN diodes. Different states of the slots will result in different radiation amplitudes and phase adjustments, which are summarized as follows:
\begin{equation}\label{eq:4}
\begin{aligned}
{P}_{k,\ell}^{+}&=
\begin{cases}
P_{0} e^{-j0} & \text{for state ON$^+$},\\
0 & \text{for state OFF$^+$},
\end{cases}\\
{P}_{k,\ell}^{-}&=
\begin{cases}
P_{0} e^{-j\pi} & \text{for state ON$^-$},\\
0 & \text{for state OFF$^-$}.
\end{cases}
\end{aligned}	
\end{equation}

Relevant to the functionality of position-flexible FAS, three states of a meta-atom are of interest: (1) OFF$^{+}$\&OFF$^{-}$; (2) ON$^{+}$\&OFF$^{-}$ and (3) OFF$^{+}$\&ON$^{-}$. This corresponds to the cases that both of the slots are off (Case 1) or either one of the slots is on (Case 2 \& Case 3), respectively, as depicted in Figure \ref{metafluid_antenna}(d). For each of the `ON' state, its magnetic field can be computed by substituting \eqref{eq:2} and \eqref{eq:4} into \eqref{eq:1}. As such, the far-field expression for the $(k,\ell)$-th meta-atom in the `ON' state can be expressed as
\begin{equation}\label{eq:5}
\begin{aligned}
{H}_{k,\ell}(\theta)=
\begin{cases}
-\left(\pi H_0 f_0^2 \cos\theta\right)  e^{j 2 \pi f_0 t}  
\left[\frac{P_{k,\ell}^{+}}{d_{k,\ell}^{+}} e^{-j\left(\xi_{\rm gw} x_{k,\ell}-k_0 d_{k,\ell}^{+} \sin \theta \right)}\right] & \text{for state ON$^{+}$\&OFF$^{-}$},\\
-\left(\pi H_0 f_0^2 \cos\theta\right)  e^{j 2 \pi f_0 t}  
\left[\frac{P_{k,\ell}^{-}}{d_{k,\ell}^{-}} e^{-j\left(\xi_{\rm gw} 	x_{k,\ell}-k_0 d_{k,\ell}^{-} \sin \theta \right)}\right] & \text{for state OFF$^{+}$\&ON$^{-}$}.
\end{cases}
\end{aligned}
\end{equation}

By switching the diodes between `ON' and `OFF' states, the radiation of the fields can be selectively blocked or allowed to pass through. A comparison of the electric field (E-field) differences between radiating and non-radiating states reveals a significant $13~{\rm dB}$ difference, as seen in Figure \ref{metafluid_antenna}(c). In the meta-fluid antenna, the slot `$+$' and slot `$-$' will not be radiating at the same time, as their radiation will cancel from each other if both slots are in the `ON' state.

\subsection*{Rich-scattering generators}
To facilitate our experiments, we need to emulate the propagation conditions and characteristics of real environments in our laboratory setup. In particular, we design a waveguide-fed generator that radiates disordered beams, resulting in diverse, random E-field patterns to mimic what appears in real rich-scattering environments. Our generator is a waveguide surface that contains many randomly distributed slots. 

To accurately determine the total surface radiation, it is necessary to account for the far-field contributions from all the slots. In this scenario, the overall far-field radiation from the generator can be characterized as
\begin{equation}\label{eq:7}
{H}(\theta)=- \left(\pi H_0 f_0^2 \cos\theta\right)  e^{j 2 \pi f_0 t} \sum_{m=1}^{M_{s}}\sum_{n=1}^{N_{s}}\left[\frac{P_{m,n}^{+}}{d_{m,n}^{+}} e^{-j\left(\xi_{\rm gw} x_{m,n}-k_0 d_{m,n}^{+} \sin \theta \right)}+\frac{P_{m,n}^{-}}{d_{m,n}^{-}} e^{-j\left(\xi_{\rm gw} x_{m,n}-k_0 d_{m,n}^{-} \sin \theta \right)}\right],
\end{equation}
where $2M_{s} \times N_{s}$ denotes the total number of slots in the waveguide structure, and the rest of the variables/parameters are defined in a similar way as in the case of meta-fluid antenna earlier, except that we now have fixed slots but not meta-atoms. Also, there are $M_s$ rows of slots and each row consists of an upper `$+$' row and a lower `$-$' row of slots. Note that in the meta-fluid antenna, the meta-atoms are tuneable but the slots here are not.

For the generator, the radiating units are discrete slots, and the input energy leaks out one after another through the slots. For this reason, we need to consider the radiation efficiency $\eta$ of each radiating unit. For the slots, their radiation efficiency is related to the attenuation constant in $y$-direction, $\alpha_y$, and the offset distance $\Delta l$ from the waveguide center, i.e.,
\begin{equation}\label{eq:8}
\eta=1-e^{-2 \alpha_y \Delta l}.
\end{equation}
Let us focus on a particular row (in the $y$-direction), i.e., any of the $M_s$ rows, upper or lower, of each generator. It is possible to work out the radiation power resulting from a particular slot, say $n_s$, which can be found as
\begin{equation}\label{eq:9}
P^{\rm slot}_{n_s}=\eta_{n_{s}} \prod_{n=1}^{n_{s}-1}\left(1-\eta_n\right) P_{\rm in},
\end{equation}
where $P_{\rm in}$ denotes the input power and $\eta_n$ can be obtained using (\ref{eq:8}). From (\ref{eq:9}), it is reasonable to see that the radiation power is stronger (or weaker) if it is closer to (or farther away from) the input. Now, if we take into account all the $M_s$ rows in the waveguide-fed generator considering the use of a power divider to split the input power onto the $M_s$ channels, then the radiation power from the $(m_s,n_s)$-th slot can be easily obtained as \cite{wu2022self}

\begin{equation}\label{eq:11}
P^{\rm slot}_{m_{s},n_{s}}=\eta_{m_{s},n_{s}} \prod_{n=1}^{n_s-1}\left(1-\eta_{m_s,n}\right) \frac{P_{\rm in}}{M_{s}},
\end{equation}
where the respective variables are extended to cope with the 2D indices on the waveguide-fed generator. Finally, substituting \eqref{eq:11} into \eqref{eq:7} gives the overall far-field radiation
\begin{multline}\label{eq:12}
{H}(\theta)=- \left(\pi H_0f_0^2 \cos\theta\right)  e^{j 2 \pi f_0 t} \\
\times\frac{P_{\rm in}}{M_{s}}\sum_{m=1}^{M_{s}}\sum_{n=1}^{N_{s}}
\left[\frac{P_{m,n}^{+}}{d_{m,n}^{+}} e^{-j\left(\xi_{\rm gw} x_{m,n}-k_0 d_{m,n}^{+} \sin \theta \right)}
+\frac{P_{m,n}^{-}}{d_{m,n}^{-}} e^{-j\left(\xi_{\rm gw} x_{m,n}-k_0 d_{m,n}^{-} \sin \theta \right)}\right]\eta_{m,n} \prod_{l=1}^{n-1}\left(1-\eta_{m,l}\right).
\end{multline}

To provide sufficient representation of real-world environments, we have made $30$ such rich-scattering generators, all different with arbitrarily placed slots. We divide them into three groups and each group represents $10$ independent wireless channel realizations from one transmitter. As a result, those generators are sufficient to test a $3$-user FAMA system over $10$ independent wireless environments. Our experiments will be discussed in the next section. Importantly, the method used to generate the environment does not affect the conclusion of this work. In fact, we find that our experimental setup yields results that more closely align with what are expected in Rayleigh fading environments. This method is therefore valid and offers robust validation of our experimental setup.

\subsection*{Experimental validation for FAMA}
\begin{figure}[!h]
\captionsetup{justification=justified, singlelinecheck=false}
\centering\includegraphics[width=17.5cm]{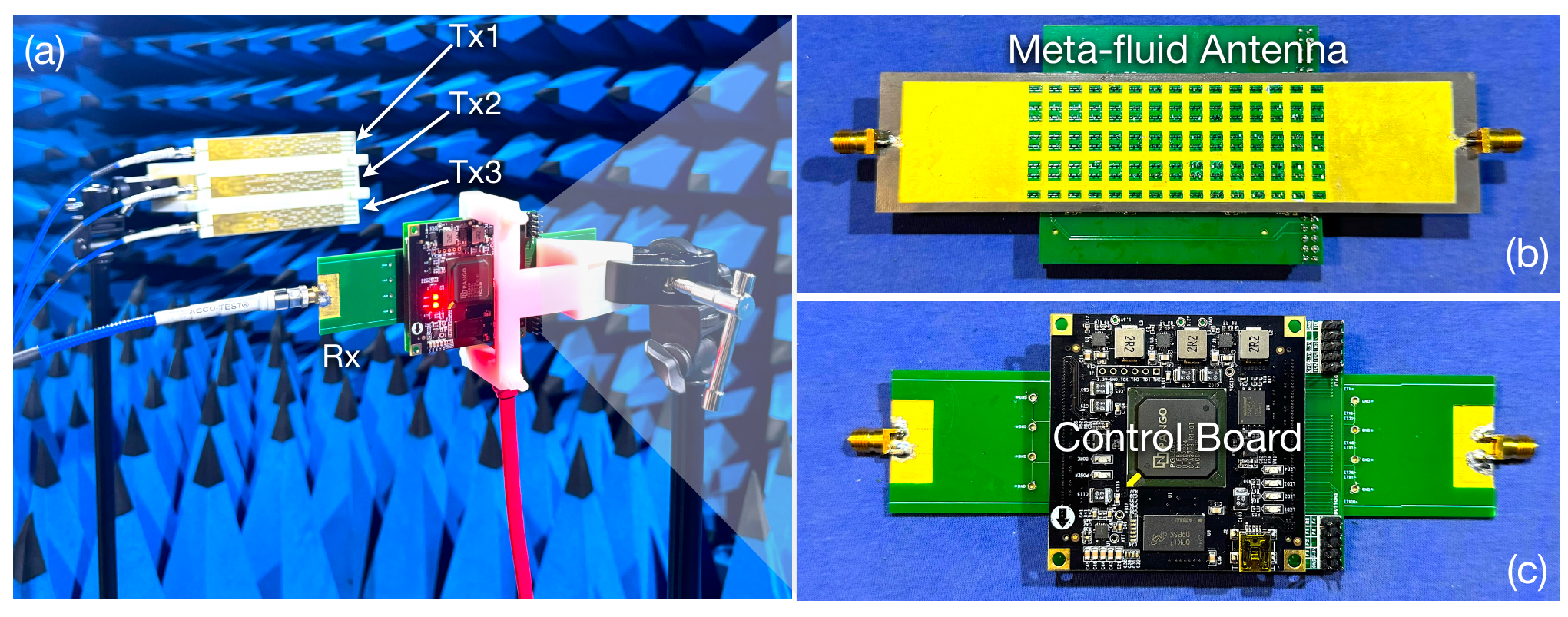}
\caption{\textbf{The experimental setup of the FAMA system utilizing the meta-fluid antenna prototype Rx.} (a) A photo shows the rich-scattering generators as the transmitter group and the meta-fluid antenna prototype in the chamber for measurement. Additionally, a red cable, which serves as the DC power cable for the FPGA, is also visible. The transmitter group and the receiver are separated by a distance of $0.5$ meters. (b) This photo shows the front view of the antenna prototype with two ports, one connecting to a $50\Omega$ load and another to the RF chain. The front side of the meta-fluid antenna is composed of $8$ rows and $15$ columns, totalling $120$ elements. Each element is controlled by $4$ PIN diodes, enabling toggling between the `radiating' state and the `non-radiating' state. (c) This photo shows the back view of the meta-fluid antenna attached to the integrated control board, which is seamlessly integrated with the antenna section through biasing vias. By reprogramming the FPGA, the radiating position of the meta-fluid antenna can be altered, allowing for dynamic control of the radiation pattern of the antenna.}\label{FAMA_system}
\end{figure}

In this section, we present a prototype of the meta-fluid antenna and the experimental setup for demonstrating the feasibility of FAMA using the proposed meta-fluid antenna prototype, as shown in Figures \ref{FAMA_system}(a). Our experiments utilize the waveguide-fed generators to emulate the rich-scattering propagating characteristics seen in real-world environments. The fabricated meta-fluid antenna is designed with a unique architecture that enables position switching, featuring $8\times 15=120$ elements, i.e., $120$ switchable positions. Each element can be independently controlled to switch between different operating states, allowing for dynamic reconfiguration of radiation position (or the resulting aperture) of the antenna. This is achieved through the use of PIN diodes, which are integrated into each element and controlled by an integrated control board. The control board is used for the operation of the meta-fluid antenna, as it connects to each PIN diode through biasing vias that provide the necessary DC current to switch the diodes between their `ON' and `OFF' states. Figures \ref{FAMA_system}(b) and \ref{FAMA_system}(c) show the front and back views of the meta-fluid antenna prototype, with the back view also showing brief details of the integrated control board. Besides, the meta-fluid antenna operates at a center frequency of $26.5~{\rm GHz}$ with a bandwidth of $1~{\rm GHz}$.

Our $30$ waveguide-fed generators represent $10$ independent experiments (Case 1 to Case 10) for a $3$-user FAMA system, with the meta-fluid antenna prototype as the receiver terminal (labelled as `Rx'). During the experiments, the meta-fluid antenna was connected to one of the four ports of a vector network analyzer (VNA), serving as the receiving antenna, while the transmitter group, comprising three random E-field emitting antennas in the form of the $3$-user waveguide-fed generators, was connected to the remaining three ports (labelled as `Tx1', `Tx2' and `Tx3', respectively). This setup enabled us to measure the received SINR of the meta-fluid antenna receiver with a varying activated position, thereby allowing us to understand the interference immunity of FAMA using the newly designed FAS. Specifically, any transmitter (Tx1, Tx2 or Tx3) can be viewed as the user of interest but if Tx1 is the user of interest, then Tx2 and Tx3 will be the interferers at the meta-fluid antenna receiver. In our experiments, we consider all the possible combinations so there is no bias towards a particular transmitter. 

\begin{figure}[!h]
\captionsetup{justification=justified, singlelinecheck=false}
\centering\includegraphics[width=17.5cm]{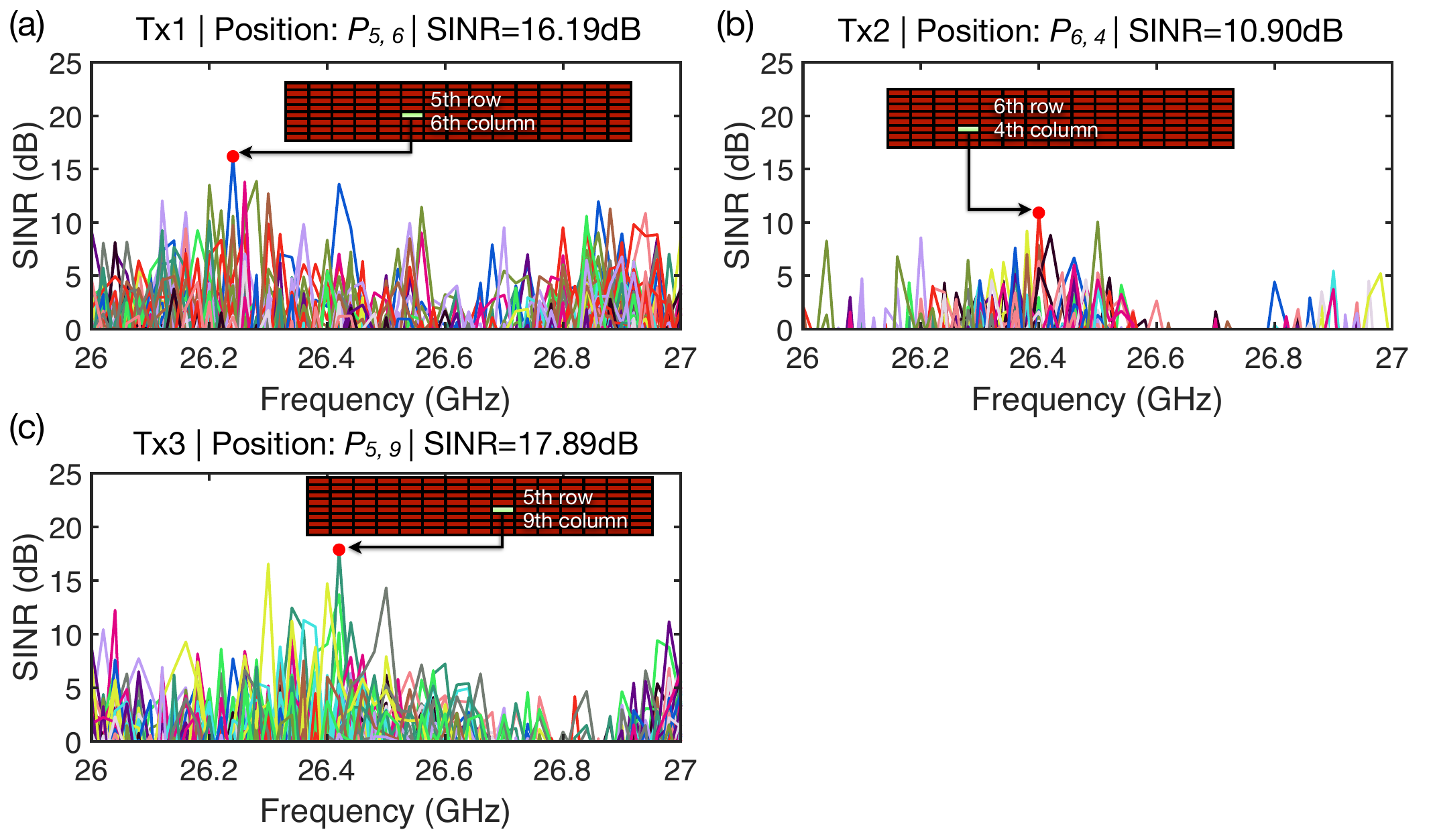}
\caption{\textbf{The SINR measurements at the meta-fluid antenna prototype in the frequency range 26 to 27 GHz in Case 9 of the transmitter group.} (a) This figure shows the measured SINRs against the frequency in the situation where Tx1 is considered as the desired user and Tx2 and Tx3 are the interferers. Moreover, at $26.25~{\rm GHz}$, the SINR at the $(5,6)$-th position of the meta-fluid antenna is highlighted, which is the maximum SINR achievable by the meta-fluid antenna ($16.19~{\rm dB}$). (b) Similar results are provided for the situation where Tx2 is the desired user and Tx1 and Tx3 are interferers. In this case, the maximum SINR at $26.4~{\rm GHz}$ is highlighted, which corresponds to $10.90~{\rm dB}$ at the $(6,4)$-th position of the meta-fluid antenna. (c) Finally, the SINR results for the situation with Tx3 being the desired user and Tx1 and Tx2 being the interferers are shown. The $(5,9)$-th position at a frequency near $26.43~{\rm GHz}$ is highlighted, achieving an SINR of $17.89~{\rm dB}$.}\label{SINR_broadband}
\end{figure}

In Figure \ref{SINR_broadband}, we provide the experimental results that illustrate the achievable SINR performance of FAMA using the meta-fluid antenna for a given channel realization (Case 5 of the generators). In Figure \ref{SINR_broadband}(a), the situation considers Tx1 being the desired user and Tx2 and Tx3 being the interferers while Figure \ref{SINR_broadband}(b) considers that Tx2 is the desired user and in Figure \ref{SINR_broadband}(c), Tx3 is the desired user, with the other transmitters being the respective interferers. The results provided include the SINR data at the meta-fluid antenna (Rx) over the frequency range between $26~{\rm GHz}$ and $27~{\rm GHz}$ and at each frequency, the received SINRs of Rx with all the $120$ possible activated positions are shown. The results clearly demonstrate that at any given frequency, the meta-fluid antenna obtains a large range of SINR values when its activated position changes, confirming the concept of FAMA via position-flexible FAS and that the benefits of FAMA are realizable by the proposed meta-fluid antenna prototype. In the figures, we have also marked the position of maximizing the SINR by FAMA at an arbitrary frequency point in all three situations, from Figures \ref{SINR_broadband}(a) to \ref{SINR_broadband}(c). It is apparent that in all three situations, the meta-fluid antenna Rx can achieve a high SINR $(>10.90~{\rm dB})$, meaning that the two interferers are satisfactorily avoided. Note that calculating the SINR of the proposed system can be accomplished by taking turn to activate the position of each meta-cell. With the high switching speed of $20~{\rm MHz}$, the SINR measurement at the FAS is easily achievable.

\begin{figure}[!h]
\captionsetup{justification=justified, singlelinecheck=false}
\centering\includegraphics[width=17.5cm]{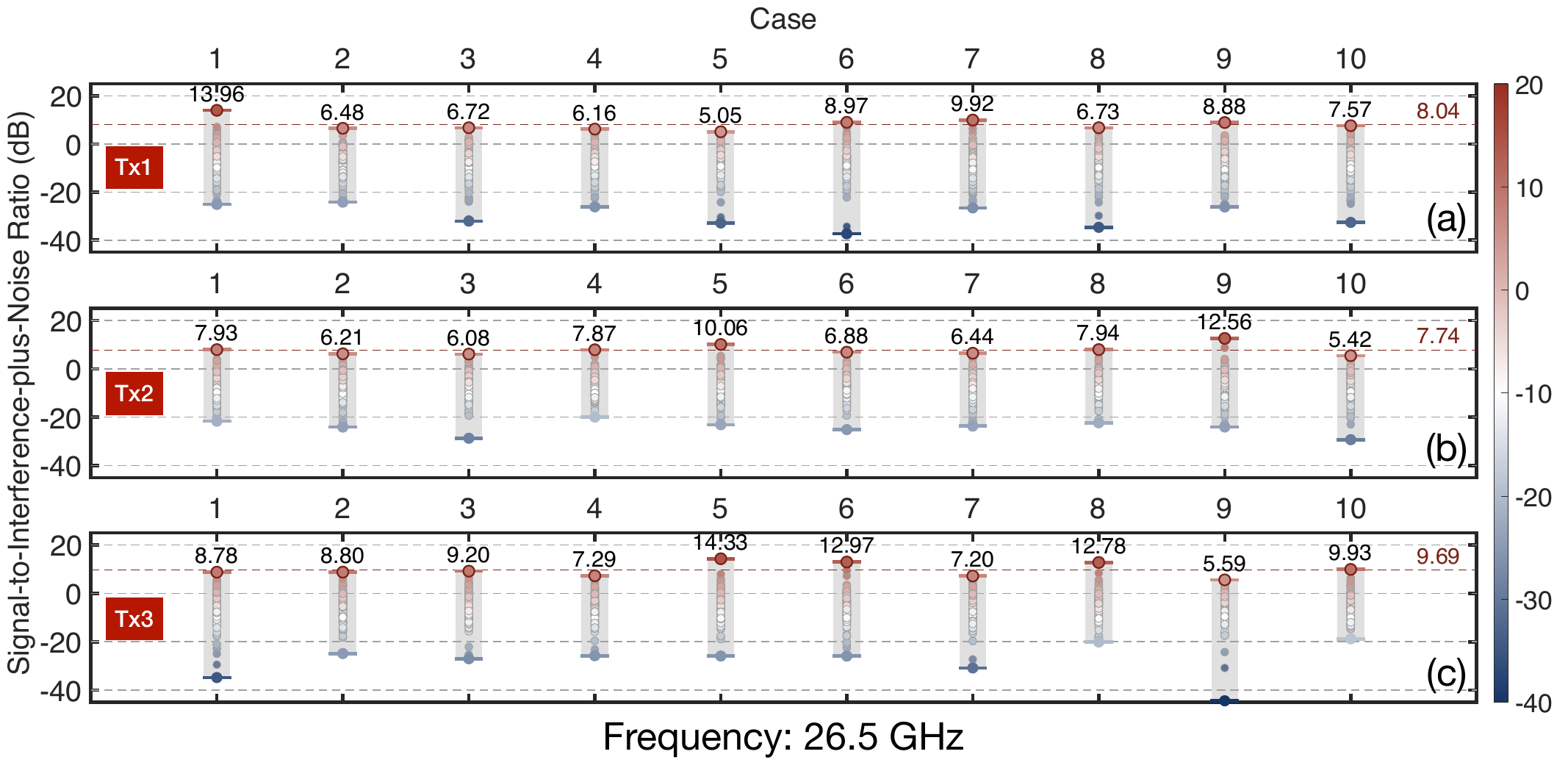}
\caption{\textbf{The SINR measurement data at the meta-fluid antenna in all the propagation situations (Case 1 to Case 10) at 26.5 GHz.} (a) This figure focuses on the situation where Tx1 is the desired user in the calculation of SINR. The ranges of the SINR observable at the meta-fluid antenna are shown in all $10$ cases. In this situation, the average SINR over all the cases is also marked as $8.04~{\rm dB}$. (b) This figure provides similar SINR data but in the situation where Tx2 is considered as the desired user. In this situation, the average SINR achievable by the meta-fluid antenna is $7.74~{\rm dB}$. (c) This figure considers the situation treating Tx3 as the desired user and Tx1 and Tx2 as the interferers. In this case, the average SINR at the meta-fluid antenna is calculated to be $9.69~{\rm dB}$.}\label{SINR_at26.5GHz}
\end{figure}

While the results in Figure \ref{SINR_broadband} are encouraging, they are limited to the Case 9 environment. Figure \ref{SINR_at26.5GHz} therefore provides the results of all the experimental environments (Case 1 to Case 10) to have a more complete understanding of the achievable performance of FAMA using the proposed meta-fluid antenna prototype. Additionally, we provide the SINR measurement data if the meta-fluid antenna Rx is replaced by a conventional fixed-position antenna for comparison. For clarity, we focus only on the results at frequency $26.5~{\rm GHz}$ but our general discussion is applicable to other operating frequencies permissible for the meta-fluid antenna. The results in Figure \ref{SINR_at26.5GHz} are given in three bar charts, each representing a particular transmitter being the desired user of interest. Ten bars illustrate the ranges of SINR obtained by the meta-fluid antenna Rx for the ten independent experiments (Case 1 to Case 10). For any given case, the results indicate a good range of SINR values, high and low, suggesting that the meta-fluid antenna experiences different levels of interference as the activated position changes--an essential feature for FAMA to work well. Evidently, in all the cases considered, the meta-fluid antenna Rx is able to achieve a decent SINR performance by simply changing the activated position. The worst case was Tx1 in Case 5 that obtained an SINR of $5.05~{\rm dB}$ while the best case was Tx3 in Case 5 achieving an SINR of  $14.33~{\rm dB}$. Note that, for our concept verification, traditional antenna metrics such as radiation pattern and S-parameters are not the focus. Instead, the objective is to confirm the ability of the proposed meta-fluid antenna to effectively avoid interference in an artificial Rayleigh fading environment. Our experimental results are confirming the successful implementation of FAS and demonstration of the FAMA concept. The choice of a horn antenna was intentional and serves as an optimistic benchmark, as it allows for better interference management and a controlled SINR. It is encouraging to see that the proposed meta-fluid antenna performs much better than this optimistic benchmark in terms of communication performance. This comparison further validates the capability of the meta-fluid antenna in rich-scattering environments.

\section*{Discussion}
It is important to understand that the proposed meta-fluid antenna does not fall into the category of conventional reconfigurable antennas even though pin-diodes are used for reconfiguration. Conventional reconfigurable antennas are limited to changing a range of radiation characteristics (e.g., operating frequency, efficiency and so on) and are never designed to have the ability to reconfigure the position of the antenna aperture. The reason is that typical antenna experts do not see the benefits of it from purely the perspective of designing a functional antenna. However, wireless communications concerns more than just the radiation efficiency of an antenna and the exact position the signal is being received is uniquely important. Specifically, a key feature to wireless communications is that the received signal at any given position is randomly attenuated (according to Rayleigh fading under rich scattering), meaning that the communication performance varies according to the antenna position at any given time. This has led to the emerging concept of FAS which has been greatly impacting the wireless communication community \cite{wong2020performance, wong2020fluid, new2024tutorial} and motivates us to design for the first time an antenna that can reconfigure the position of its aperture to exploit the signal variation in the spatial domain. Our main contribution is the meta-fluid antenna architecture that can dynamically change the position of its radiation aperture over a 2-dimensional (2D) designated area by controlling the pin-diodes. This adaptability leverages the spatial diversity within a Rayleigh fading environment, allowing us to exploit channel variations for enhanced signal reception and mitigate interference. Moreover, this work represents the first experimental validation of the FAMA concept achieved through a new form of reconfigurable antenna, referred to as FAS, a contribution we believe is unprecedented in existing literature.

Contrasting to MIMO in which its spatial diversity is dictated by the number of RF chains at the mobile side, FAS unleashes a totally new degree of freedom through position reconfigurability in the spatial domain even when the number of RF chains is fixed. Specific to multiuser communications on the same physical channel, FAS recognizes the classical fading phenomenon that signal magnitude fades differently at different locations depending upon how multiple propagation paths combine, and uses it to our advantage for avoiding interference at the receiver end by adapting the antenna position. This fundamentally changes the way in which interference is mitigated in wireless networks. Conventionally, this is done by complex signal processing (called precoding) at the BS side which requires the availability of CSI of all users involved. The FAS approach referred to as FAMA, however, eliminates the need of CSI estimation and feedback to the transmitter as well as the complex optimization of precoding, and deals with interference as they arrive at the receiver. Therefore, there is great potential of FAMA to improve the scalability of multiuser wireless communications. While theoretical results reporting the amazing performance of FAMA have emerged in recent years and it is always desirable to consider a wide range of performance metrics for a more complete evaluation, experimental results are absent and there is also no known prototype successfully designed for FAS and for validation of FAMA. This work is therefore original and significant in that this serves as the first-ever FAS antenna design realizing the concept of position reconfigurability, and tested experimentally for FAMA.

Finally, it is worth mentioning that CSI estimation for FAS has been addressed in \cite{xu-2024} and machine learning techniques have also been shown to find many applications on different FAS communication problems, e.g., \cite{chai2022port,Waqar-2024cl,wong2024virtual,Zhang-2024,Wang-2024isac,Waqar-2024ofamma,Eskandari-2024,Wang-2024wc}. For more discussion, readers are referred to \cite{new2024tutorial}. In future work, it would be important to extend the experimental setup to include real propagation environments. While our current setup, using artificial rich-scattering generators, effectively emulates Rayleigh fading conditions, testing in real environments will provide stronger evidence of FAS capabilities and the robustness of FAMA in realistic scenarios.

\section*{Conclusion}
While the proposition of mitigating interference by antenna position reconfigurability following the emerging concept of FAS is exciting and numerous theoretical work has already reported its extraordinary performance to wireless communications, experimental validation is lacking and the effectiveness of FAS remains a hopeful speculation. To fill this gap, this paper has made two key contributions. First, by leveraging programmable waveguide feeding, we designed and fabricated a software-controlled meta-fluid antenna that can reconfigure the position of radiation on demand over a given surface, achieving the functionality of a position-flexible FAS for the first time. Figuratively speaking, the proposed meta-fluid antenna acts like a genie riding on radio wave fluctuation created by nature in the spatial domain, navigating and activating the point in space where the communication condition becomes desirable. This is the unique feature of FAS which eliminates the need of expensive signal processing at the transmitter side, a major obstacle in the state-of-art precoding scheme used in 5G. The second major contribution of this paper is to provide experimental validation of using FAS for multiuser communications on a shared channel, referred to as FAMA in the communication community. Specifically, we fabricated a large number of waveguide-fed generators with arbitrarily placed slots to emulate real-world radio propagation environments with rich scattering. The idea of FAMA was put to test using the prototype as the receiver terminal in a $3$-user scenario (one desired transmitter and two interferers). Our experiments validated that FAMA is feasible using the proposed meta-fluid antenna and an average of more than $15~{\rm dB}$ received SINR is obtained over all the cases considered. Overall, our findings provided strong experimental evidence that FAMA is practically achievable and obtains promising performance that paves the way for scalable multiple access technologies. Last but not least, the results demonstrated the superiority of the proposed meta-fluid antenna over a conventional fixed-position antenna, reaffirming the importance of FAS.

\end{document}